\documentclass{elsart}

\usepackage{graphicx}
\usepackage{amssymb}

\begin{document}

\begin{frontmatter}



\title{First Scanning Tunnelling Spectroscopy on Bi$_2$Sr$_2$Ca$_2$Cu$_3$O$_{10+\delta}$ single crystals}
\author{M.~Kugler},
\author{G.~Levy~de~Castro},
\author{E.~Giannini},
\author{A.~Piriou},
\author{A.~A.~Manuel},
\author{C.~Hess} and
\author{{\O}.~Fischer}
\address{DPMC, Universit\'e de Gen\`eve, 24, Quai Ernest-Ansermet, 1211 Gen\`eve, Switzerland}

\begin{abstract}
We report the first low temperature scanning tunnelling microscopy
and spectroscopy study of high quality
Bi$_2$Sr$_2$Ca$_2$Cu$_3$O$_{10+\delta}$ crystals. We present
atomic resolution and show spectroscopic data acquired on two
different samples. In one case, for $T_c=109$~K and a transition
width of only 1~K, we obtained an extremely homogeneous sample
with $\overline{\Delta}_p=60$~meV over at least 50~nm. In the
other case, the respective parameters were $T_c=111$~K and $\Delta
T_c=1.7$~K and yielded a slightly less homogeneous sample with
$\overline{\Delta}_p=45$~meV. We evidence strong similarities with
Bi$_2$Sr$_2$CaCu$_2$O$_{8+\delta}$ and discuss the doping level of
our samples.
\end{abstract}

\end{frontmatter}

% main text
\section{Introduction}
%\label{}

The trilayer cuprate
Bi$_2$Sr$_2$Ca$_2$Cu$_3$O$_{10+\delta}$~(Bi2223) attracted a
strong interest due to its high critical temperature
$T_c^{max}=111$~K and its potential for applications. However, the
difficulty in synthesizing large sized single-phase crystals
significantly impeded the study of fundamental properties. Very
recently the effort in developing new crystal growth processes was
rewarded by a successful production of homogeneous high quality
Bi2223 single crystals~\cite{Fujii-2001}, thus opening the door to
scanning tunnelling spectroscopy~(STS) investigations. The
determination of the intrinsic superconducting properties of this
trilayer compound is of crucial importance for the determination
of the generic features and behaviors of Bismuth based cuprates
and more generally in the quest of the understanding of high-$T_c$
superconductivity.

In this paper, we present the first low temperature STS study of
high quality Bi2223 single crystals. We investigate the spatial
dependence of the local density of states~(LDOS) of two different
samples and discuss parallels with the parent double layer
compound Bi$_2$Sr$_2$CaCu$_2$O$_{8+\delta}$~(Bi2212).

\section{Sample characterization}
%\label{}

We studied two Bi2223 single crystals grown by the travelling solvent
floating zone method~\cite{Giannini-2004}, which were
post-annealed under different conditions. Sample $A$ was treated
at 500~$^{\circ}$C, 20~bar O$_2$ during 50~hours and yielded
$T_c^{onset}=109$~K and an extremely sharp transition width
$\Delta T=1$~K as measured by SQUID magnetometry. For sample $B$
the conditions were 500~$^{\circ}$C, 100~bar O$_2$ during
50~hours, giving $T_c^{onset}=111$~K and a larger transition width
$\Delta T=1.7$~K, indicating less homogeneity. The measurements
presented here were obtained using a home built scanning
tunnelling microscope which operates under ultrahigh vacuum and
low temperatures~\cite{Kugler-2000a}. The vacuum tunnel junctions
were made between the in-situ cleaved Bi2223 (001) surface and an
electrochemically etched iridium tip mounted perpendicular to the
sample surface. All spectra where obtained at 1.8~K and zero
magnetic field with tunnelling resistances ranging between 0.5 and
1~G$\Omega$.

\section{Results and discussion}
%\label{}

\begin{figure}
\includegraphics [width=7cm,clip] {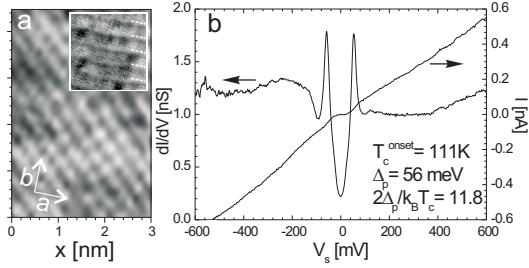}
\caption{(a) Topographic image of in-situ cleaved Bi2223 obtained
under UHV at 1.8K ($I$=0.8nA, $V_s$=0.6V). The 13x13nm$^2$ inset shows the 26~{\AA} supermodulation. (b) Typical $IV$-curve and
corresponding $dI/dV$ spectrum acquired on sample $B$ at 1.8K
($R_t$=1G$\Omega$).} \label{fig1}
\end{figure}

In Fig.~\ref{fig1}a we present a topographic scan showing atomic
resolution. We further observe a 26~{\AA} supermodulation along
the $b$-axis which is also seen in Bi2201~\cite{Kugler-2001} and
Bi2212~\cite{Renner-1995} where it is attributed to extra oxygen
in the BiO planes~\cite{Zandbergen-1998}. In Fig.~\ref{fig1}b we
show a typical $IV$-curve and its corresponding differential
conductance. The IV-characteristic clearly shows a metallic-like
background over a large energy range and the existence of a clear
$d$-wave gap at the Fermi level. The differential tunnelling
conductance presents well-developed coherence peaks at
$\pm\Delta_p\simeq56$~meV and clear dip-hump structures which
appear asymmetrically on both sides of the Fermi level. These
features are consistent with
Bi2212~\cite{Renner-1995,Zasadzinski-2001}, where the broad hump
observed at negative bias has been related to a van Hove
singularity~\cite{Hoogenboom-2003b}.

\begin{figure}
\includegraphics [width=7cm,clip] {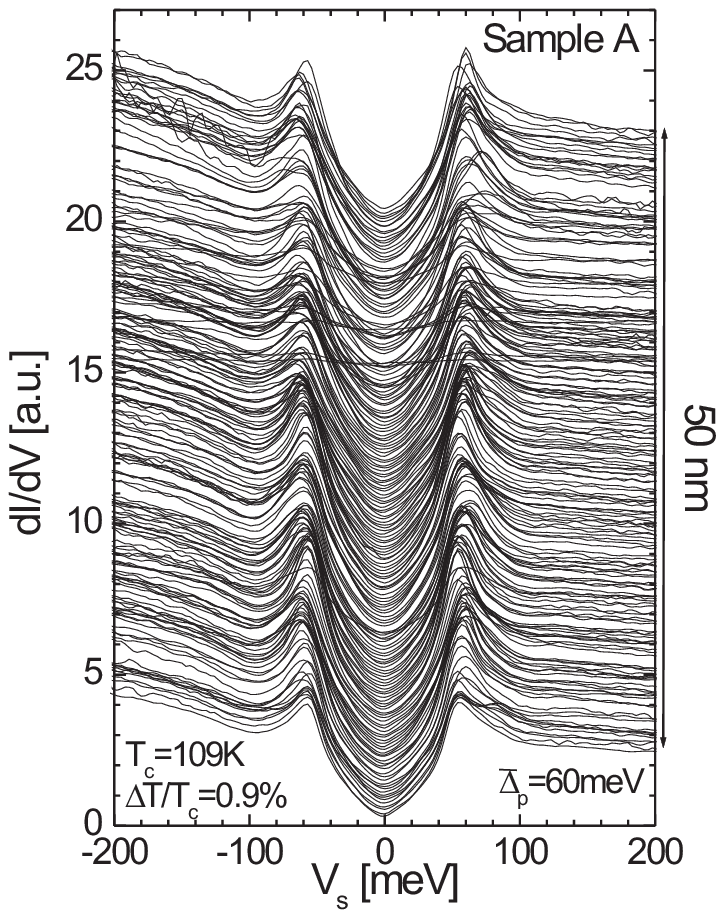}
\caption{Trace of 201 spectra along 50nm line on sample $A$ at
1.8K ($I$=0.6nA, $V_s$=0.3V).} \label{fig2}
\end{figure}

\begin{figure}
\includegraphics [width=7cm,clip] {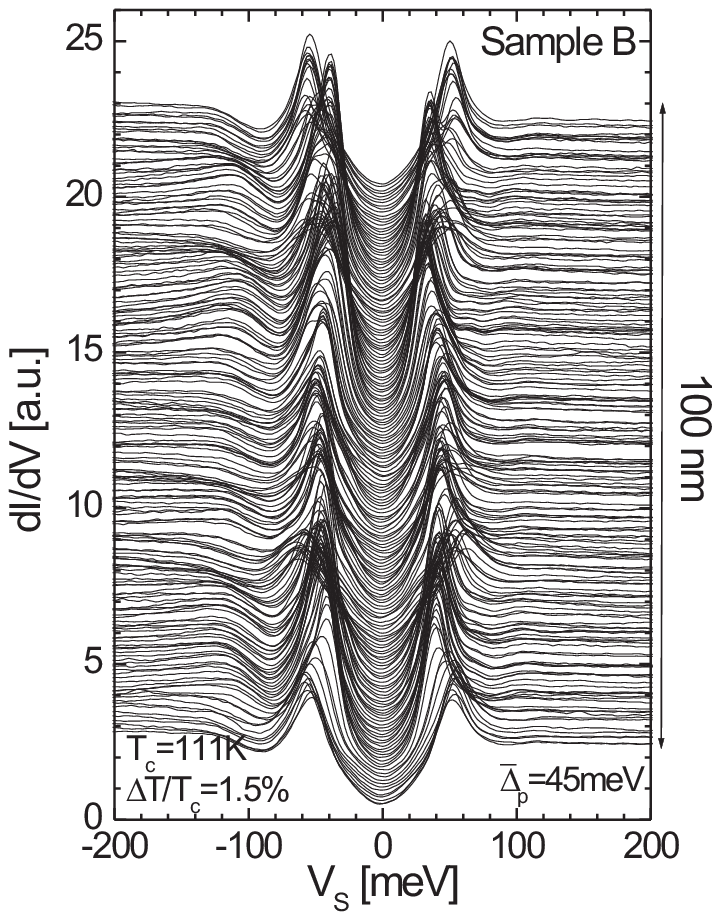}
\caption{Trace of 201 spectra along 100nm line on sample $B$ at
1.8K ($I$=0.6nA, $V_s$=0.6V).} \label{fig3}
\end{figure}
The power of STS lies in the ability to investigate the spatial
dependence of the LDOS. Recently, much attention has been paid to
the question whether the electronic structure of high-$T_c$
cuprates is intrinsically inhomogeneous at the nanometer
scale~\cite{Lang-2002} or if long range homogeneity of the
spectral signature can be obtained. In fact, for overdoped Bi2212
it has been demonstrated that an appropriate crystal preparation
yields homogeneous samples~\cite{Hoogenboom-2003a}. A necessary,
however not exclusive, criterium for sample homogeneity is a sharp
superconducting transition width $\Delta T/T_c<1\%$ in
susceptibility measurements. 
\begin{figure}
\includegraphics [width=7cm,clip] {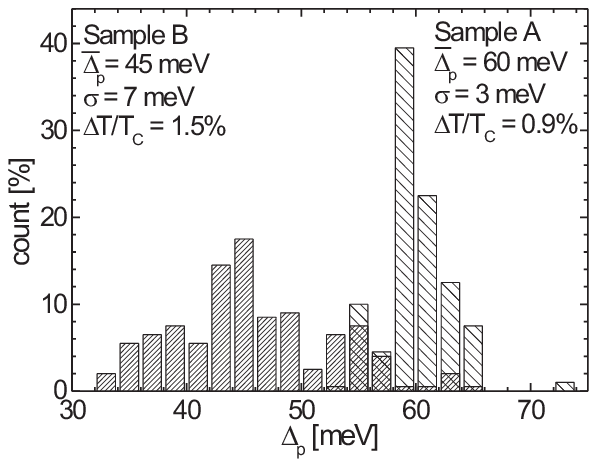}
\caption{Gap distributions of sample $A$ and $B$.} \label{fig4}
\end{figure}

\begin{figure}
\includegraphics [width=7cm,clip] {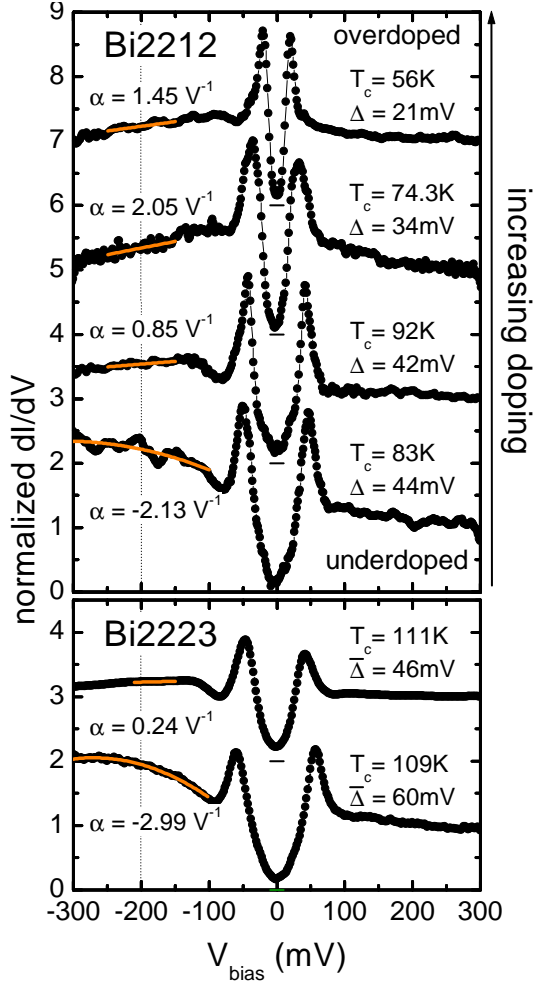}
\caption{Comparison of the LDOS doping dependence between Bi2212
(Ref.~\cite{Renner-1998a}) and Bi2223. All spectra have been
normalized to the background conductance at $V_s$=300meV and
shifted vertically for clarity. $\alpha$ is the slope of the
background fit at $V_s$=-200meV. The Bi2223 panel shows the
average spectra of the Figs.~\ref{fig2} (bottom)
and~\ref{fig3} (top).} \label{fig5}
\end{figure}

\begin{figure}
\includegraphics [width=7cm,clip] {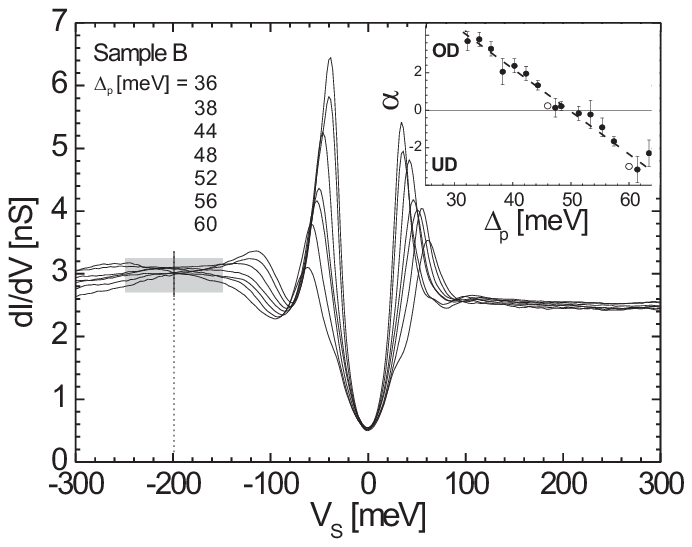}
\caption{Average spectra corresponding to energy intervalls of the
gap distribution of sample $B$ (see Fig.~\ref{fig4}). The shaded
energy intervall indicates where the background has been linearly
fitted to determine the slope $\alpha$. The inset shows the
relation between the background slope at $V_s$=-200meV and
$\Delta_p$. Open symbols correspond to the average spectra shown
in the bottom of Fig.~\ref{fig5}.} \label{fig6}
\end{figure}
In the Figs.~\ref{fig2}
and~\ref{fig3} we compare traces of equidistant spectra acquired
along a line on the surfaces of sample $A$ and $B$ respectively.
As anticipated by the sharp transition width $\Delta T/T_c=0.9\%$,
sample $A$ shows an extremely high homogeneity, as demonstrated by
the gap distribution in Fig.~\ref{fig4}: the standard deviation
$\sigma$ is of only 3~meV for a mean gap
$\overline{\Delta}_p=60$~meV. In contrast, sample $B$, with
$\Delta T/T_c=1.5\%$, shows a much broader distribution
$\sigma=7$~meV and $\overline{\Delta}_p=45$~meV. This behavior is
consistent with the empirical criterium mentioned above and
demonstrates that homogeneity can be obtained on Bi2223 over a
range of at least 50~nm. Note the spatial reproducibility of the
background conductance outside the peak-dip-hump structure for
both traces, which is characteristic for a stable vacuum
tunnelling junction.

The doping relation $T_c(p)$ of Bi2223 has not been established
yet. However, various observations point out that $T_c(p)$ could
deviate from the generic dome shaped relation by Presland {\it et
al.}~\cite{Presland-1991}. Indeed, the doping dependence has been
studied by other methods suggesting a large plateau with
$T_c=T_{c, max}$ in the overdoped
phase~\cite{Fujii-2002,Yamada-2003a}. Furthermore, it has been
found that Bi2223 is far less sensitive to oxygen doping than
Bi2212 since upon post-annealing only small variations of $T_c$
are achieved, even with high oxygen pressure
treatments~\cite{Fujii-2002}. With that respect it is striking
that for a variation of only $2\%$ of $T_c$ we observe a gap
variation of about $28\%$ between sample $A$ and $B$.

Two observations allow us to tentatively assign sample $A$ and $B$
as being underdoped and optimally doped, respectively. Firstly,
the gap magnitude is larger in sample $A$ than in $B$. Assuming
that the gap falls monotonically with increasing doping, as
established for Bi2212~\cite{Renner-1998a,Miyakawa-1998} and
suggested for Bi2223 by $c$-axis conductance
experiments~\cite{Yamada-2003a}, this indicates that sample $A$ has
a lower hole content than $B$. Secondly, for sample $B$ we have
$T_c=T_{c,max}$, which is characteristic for optimally or possibly
overdoped samples~\cite{Fujii-2002}.

This conclusion is corroborated by a careful comparison of the
background conductance in Bi2212 and Bi2223 spectra. Various
tunneling experiments on Bi2212 revealed that the background
conductance outside the gap is asymmetric and varies with
doping~\cite{Renner-1998a,Zasadzinski-2001}, the strongest effect
occuring below the Fermi level. The background slope $\alpha$
taken reasonably far from the Fermi level and the dip-hump
structure ($V_s\simeq -200$~meV) is positive for overdoped
samples, about zero for optimal doping and negative for the
underdoped case, as illustrated in the top panel of
Fig.~\ref{fig5}. In Bi2212, this observation has been attributed
to a broad van Hove singularity at negative bias that shifts away
from the Fermi level as $p$ decreases~\cite{Hoogenboom-2003b}. In
the lower panel of Fig.~\ref{fig5} we show the average spectra of
the traces on sample $A$ and $B$. Focusing on the background slope
at negative bias shows a striking similarity between underdoped
Bi2212 and sample $A$, and respectively between optimally doped
Bi2212 and sample $B$.

We now turn to the analysis of the LDOS variations observed along
the trace of sample $B$. Fig.~\ref{fig6} displays the average
spectra corresponding to the energy intervalls of the gap
distribution shown in Fig.~\ref{fig4}. The systematic decrease of
the coherence peak intensity with increasing gap magnitude and the
corresponding dip-hump shift, are strikingly similar to what is
observed in Bi2212 when reducing the
doping~\cite{Matsuda-2003,McElroy-2004a}. Furthermore, for each
spectrum we again determined the slope $\alpha$ at -200~meV which
appears to depend lineraly on the gap magnitude (see
Fig.~\ref{fig6}~inset). In addition the sign change of the slope
occurs at what is believed to be optimal doping. With the
tentative background versus doping relation discussed above, it is
therefore possible to interprete the spatially varying LDOS in
less homogeneous samples as being related to a local variation of
doping, as reported in Bi2212~\cite{Pan-2001}.

\section{Conclusion}
%\label{}

We showed the first STS study on Bi2223 single crystals. Atomic resolution as well as
the characteristic supermodulation have been observed. One sample with a transition width
of only $\Delta T/T_c=0.9\%$ exhibits an extremely homogeneous LDOS.
The LDOS features, i.e. the $d$-wave spectral shape, the strong
coherence peaks and the existence of a dip-hump structure, as well
as the various LDOS on the surface of a less homogeneous
sample, show striking similarity with Bi2212. Using the doping
dependence of the background conductance and of the gap of Bi2212
as a template, we made a qualitative estimate of the doping level
of our samples. We further evidenced a linear dependence and sign
change of the slope of the background as a function of the gap
magnitude.

\section{Aknowledgements}
%\label{}

This work was supported by the Swiss National Science Foundation
through the National Centre of Competence in Research "Materials
with Novel Electronic Properties - MaNEP". C.H. acknowledges support by the Deutsche Forschungsgemeinschaft through grant HE3439.

%\bibliographystyle{prsty}
%\bibliography{Bi2223paper}
%%begin{thebibliography}{00}


% notes:

% \bibitem{label2}
% If there is a note, it should come last:
% \bibitem{label3} \note
% \end{subbibitems}

%%\bibitem{}

%%\end{thebibliography}

\end{document}